\documentclass{article}
\usepackage{amsmath}
\usepackage{amssymb}
\usepackage{graphicx}
\usepackage{hyperref}

\newcommand{\D}{\,\mathrm{d}}
\newcommand{\E}{\mathrm{e}}
\newcommand{\I}{\mathrm{i}}
\newcommand{\pd}[2]{\ensuremath{\frac{\partial #1}{\partial #2}}}
\newcommand{\avrg}[1]{\ensuremath{\left<#1\right>}}
\newcommand{\n}[1]{\hat{n}_\mathbf{#1}}
\newcommand{\g}[1]{\hat{\gamma}_\mathbf{#1}}
\renewcommand{\a}[1]{\hat{a}^{\phantom{+}}_\mathbf{#1}}
\renewcommand{\aa}[1]{\hat{a}^\dagger_\mathbf{#1}}

\date{}

\begin{document}

\title{Squeezed Vibrational States in Superfluid Helium}
\author{L.A.\,Melnikovsky$\,^{a,b}$
\\
$^a\,$Weizmann Institute of Science, Rehovot, Israel\\
$^b\,$P.L.\,Kapitza Institute for Physical Problems, Moscow, Russia
}

\maketitle

\begin{abstract}
Ultrafast birefringence oscillations observed in superfluid helium provide evidence for anisotropic quantum squeezing of quasiparticle pairs. The measured response is a superposition of contributions from all vibrational modes, with dominant contributions from rotons, maxons, and Pitaevskii's plateau. The nonzero initial phase follows naturally from multimode interference.
\end{abstract}

\section{Introduction}

Milner, Stamp, and Milner \cite{exp} recently discovered ultrafast dynamics in superfluid helium triggered by an intense femtosecond laser pulse. In their experiment, bulk liquid $^4$He below the lambda point was illuminated by focused linearly polarized pump pulses with peak intensity $I \sim 10^{12}\,\text{W}/\text{cm}^2$ and duration $\tau \sim 70\,\text{fs}$. These pulses induce oscillating optical birefringence, detected by measuring the polarization change of a time-delayed probe pulse. The oscillation frequency corresponds to twice the roton energy, $2\Delta_R \sim 17\,\text{K}$, and was interpreted as excitation of a dense cloud of roton pairs through stimulated Raman scattering. A notable unresolved feature is the nonzero initial phase of the oscillatory signal \cite{milner_private}.

Birefringence in (intrinsically isotropic) superfluid helium can be generated by an anisotropic state of elementary excitations~\cite{AF}: quasiparticles are local density fluctuations that modulate the dielectric constant; in an anisotropic population these fluctuations give rise to a non-scalar part of the macroscopic permittivity tensor. The contribution of each excitation can be conveniently parameterized by a polarizability $\alpha$ --- the same quantity that governs the quasiparticle energy shift $-\alpha E^2/2$ in an external electric field $E$. Roton polarizability is estimated~\cite{rpr} to be $\alpha_R \sim 4.5\cdot 10^{-26}\,\text{cm}^3$, which corresponds to the energy shift in the field of the pump pulse $2\pi \alpha_R I / c \sim 0.7\,\text{K}$. This shift is smaller than the roton gap $\Delta_R$ itself, but (depending on the sign of $\alpha_R$) further increase of the pump intensity $I$ may lead to more complicated effects such as superfluid breakdown, condensation of rotons, or solid nucleation~\cite{ghost}.

\begin{figure}[ht]
    \centering
    \includegraphics{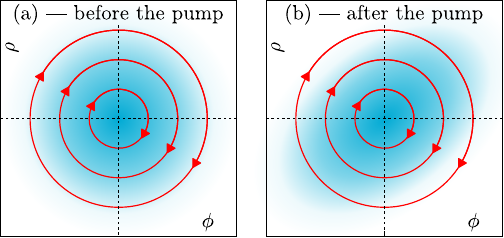}
    \caption{Wigner quasi-probability distribution in the phase space $(\phi, \rho)$. The thermal state in (a) is stationary under the circular flow of free harmonic oscillator evolution (shown in red). The pump pulse ``skews'' the distribution such that it is no longer circularly symmetric. Birefringence, given by the average $\avrg{\rho^2}$, is unchanged by the pump itself, but oscillates subsequently since the squeezed state (b) is no longer stationary.}
    \label{gauss}
\end{figure}

This paper aims to identify the state produced by the pump pulse and to explain the observed dynamics. Parametric excitation of rotons by periodic electric fields has been discussed before~\cite{rpr}. Here we show that the same coupling naturally produces an anisotropic squeezed quasiparticle state~\cite{coherent}, giving rise to the observed birefringence oscillations. Similar squeezing phenomena for phonons in solids have been investigated theoretically and observed experimentally~\cite{HuNori1996,GarrettMerlin1997}.

The suggested mechanism is illustrated in Fig.\,\ref{gauss} for a single long-wavelength vibrational mode (standing sound wave) treated as a harmonic oscillator in phase space $(\phi,\rho)$, with $\phi$ and $\rho$ denoting the corresponding Fourier components of velocity potential and density~\cite{LL9}. The thermal state is stationary under free evolution (circular flow if normalized properly). The pump electric field transiently changes the quadratic Hamiltonian by effectively modifying the ``force constant'', so that after the pulse the state becomes squeezed. Note that squeezing here refers to the quantum state of each individual mode in its own phase space, not to anisotropy in real three-dimensional space. This state is essentially non-classical and oscillates at twice the oscillator frequency. The experimental signal is the multimode sum of such contributions, which explains the nonzero initial phase.

\section{Dielectric properties}
The pump electric field $\mathbf{E}$ can be treated classically. It modifies the superfluid Hamiltonian as follows:
\begin{equation}
    \hat{H}=\hat{H}_0 - V \frac{E_m E_n}{8\pi} \hat{\epsilon}^{mn},
\end{equation}
where $V$ is the interaction volume, $\hat{\epsilon}^{mn}$ is the permittivity tensor operator, and
\begin{equation*}
\hat{H}_0=
\sum_{\mathbf{k}}
\varepsilon_k
\left(\aa{k}\a{k}+\frac{1}{2}\right)
\equiv
\sum_{\mathbf{k}}
\varepsilon_k
\left(\n{k}+\frac{1}{2}\right)
\end{equation*}
is the unperturbed Hamiltonian expressed in terms of quasiparticle creation and annihilation operators $\aa{k}$ and $\a{k}$ for each wavevector $\mathbf{k}$.

The non-scalar part of the permittivity tensor is given by~\cite{AF}:
\begin{equation}
\label{epsilon0}
\delta \epsilon^{mn}=
-\frac{1}{\epsilon} \left(\pd{\epsilon}{\rho}\right)^2
\frac{1}{V}\sum_{\mathbf{k}}
\frac{k^m k^n}{k^2}
\rho_\mathbf{k} \rho_\mathbf{-k},
\end{equation}
with $\epsilon$ the dielectric constant, $\rho$ the unperturbed fluid density, and $\rho_\mathbf{k}$ the Fourier amplitude of the density fluctuation. The scalar part does not affect birefringence and is ignored in what follows. To obtain the operator form $\hat{\epsilon}^{mn}$, we substitute the corresponding operator expression for $\rho_\mathbf{k}$, setting $\hbar=1$ here and below:
\begin{equation*}
\hat{\rho}_\mathbf{k}=
\left(\frac{\rho k^2}{2\varepsilon_k}\right)^{1/2}
\left(\a{k}+\aa{-k}\right).
\end{equation*}
This leads to
\begin{equation}
\label{epsilon1}
\hat{\epsilon}^{mn}=
\frac{4\pi}{V}
\sum_{\mathbf{k}}
\alpha^{mn}_\mathbf{k}
\left(\n{k} + \frac{\a{k}\a{-k} + \aa{k}\aa{-k} + 1}{2}\right)
,
\end{equation}
which parallels the usual permittivity formula for a gas of polarizable particles. The single-quasiparticle polarizability is defined by
\begin{equation}
\label{alpha}
\alpha^{mn}_\mathbf{k}=
-
\left(\pd{\epsilon}{\rho}\right)^2
\frac{\rho}{4\pi \epsilon}
\frac{k^m k^n}{\varepsilon_k}
.
\end{equation}
Equation \eqref{epsilon0} is valid only in the long-wavelength acoustic approximation, \eqref{alpha} is limited to the same regime. For higher momentum excitations, the most general relation is
\begin{equation}
\label{alpha-general}
\alpha^{mn}_\mathbf{k}=
\alpha_k \frac{k^m k^n}{k^2},
\end{equation}
where the phenomenological scalar $\alpha_k$ depends only on the wavevector magnitude $k$. Equation \eqref{alpha} can be used to obtain a convenient approximation for $\alpha_k$:
\begin{equation}
\alpha_k 
\sim
-
\frac{(\epsilon-1)^2 k^2}{4\pi \rho \varepsilon_k}.
\end{equation}

As in the Bogoliubov description of a weakly interacting Bose gas, \eqref{epsilon1} is non-diagonal in the quasiparticle number basis and mixes opposite-momentum states. It is therefore convenient to rewrite the sum over momentum half-space:
\begin{equation}
\label{epsilon2}
\hat{\epsilon}^{mn}=
\frac{4\pi}{V}
\sum_{\mathbf{k}>0}
\alpha^{mn}_\mathbf{k}
\g{\pm\mathbf{k}},
\end{equation}
with
\begin{equation}
    \label{gamma1}
    \g{\pm\mathbf{k}}=
    \n{k} + \n{-k} + \a{k}\a{-k} + \aa{k}\aa{-k}.
\end{equation}
We omit additive constants from operator ordering throughout, since they do not influence measurable dynamics.

The Hamiltonian likewise decomposes into independent terms from each $\pm\mathbf{k}$ pair:
\begin{equation*}
\hat{H}=\sum_{\mathbf{k}>0} \hat{h}_{\pm\mathbf{k}},
\end{equation*}
where (suppressing subscript $\mathbf{k}$)
\begin{equation}
\label{h-definition}
\hat{h}=
\left(\varepsilon+\zeta\right) \left(\n{+}+\n{-}\right)
+
\zeta\left(\a{+}\a{-} + \aa{+}\aa{-}\right).
\end{equation}
The parameter
\begin{equation*}
\zeta = -\frac{E_m E_n \alpha^{mn}_\mathbf{k}}{2}
\end{equation*}
is the leading-order quasiparticle energy shift induced by the electric field, justifying the identification of $\alpha^{mn}_\mathbf{k}$ as the quasiparticle polarizability.

\section{Squeezed state dynamics}
For a rectangular pump pulse of duration $\tau$ with linear polarization, the pump field enters as a constant RMS value because superfluid time scales are much longer than inverse optical frequency. Starting from thermal equilibrium, the pump evolution operator $\exp(-\I\hat{H}\tau)$ excites a squeezed quasiparticle state, which evolves freely under $\exp(-\I\hat{H}_0 t)$ afterwards. The contribution of each $\pm\mathbf{k}$ pair to the permittivity is the thermal average
\begin{equation}
\gamma(t)=
\avrg{
    \E^{\I \hat{h} \tau}\, \E^{\I \hat{H}_0 t}\,
    \g{}\,
    \E^{-\I \hat{H}_0 t}\, \E^{-\I \hat{h} \tau}
}.
\end{equation}
Only the terms $\a{+}\a{-}$ and $\aa{+}\aa{-}$ in \eqref{gamma1} contribute to the oscillating part of the signal:
\begin{equation}
\label{gamma2}
\gamma(t)=
\avrg{
    \E^{-2\I\varepsilon t}
    \E^{\I \hat{h} \tau}\,
    \a{+}\a{-}
    \,
    \E^{-\I \hat{h} \tau}
     + \text{h.c.}
}.
\end{equation}
We defer exact evaluation of the operators
\begin{equation}
    \label{operators}
\begin{aligned}
\hat{A}_+ &=  \E^{\I \hat{h} \tau}
                        \,\a{+}\,
                    \E^{-\I \hat{h} \tau},\\
\hat{A}_- &=  \E^{\I \hat{h} \tau}
                        \,\a{-}\,
                    \E^{-\I \hat{h} \tau}
\end{aligned}
\end{equation}
to the Appendix, but in the experiment~\cite{exp} the pump pulse is very short $\zeta\tau < \varepsilon\tau \ll 1$, and the first-order expansion in $\tau$ of the expressions above is sufficient:
\begin{equation}
\label{operators-approx}
\begin{aligned}
\hat{A}_+ & \approx
  (1-\I \tau (\varepsilon + \zeta))\a{+} - \I\tau \zeta\aa{-}
  ,\\
\hat{A}_- &\approx
  (1-\I \tau (\varepsilon + \zeta))\a{-} - \I\tau \zeta\aa{+}.
\end{aligned}
\end{equation}
Substituting these expressions into \eqref{gamma2}, we get
\begin{equation}
\label{gamma3}
\gamma_\mathbf{\pm k}(t)
\approx
-2 \zeta\tau
    \left(n_{\mathbf{k}} + n_{-\mathbf{k}} +1\right)
    \sin 2\varepsilon t.
\end{equation}

\section{Mode interference}
The pump pulse simultaneously squeezes all modes with different frequencies. The measured signal is a superposition \eqref{epsilon2} of contributions \eqref{gamma3} from all modes. Using isotropy of the quasiparticle distribution $n_\mathbf{k}$ and integrating over angles, we get
\begin{equation}
\label{result}
\epsilon^{mn}=
\frac{4\tau E_m E_n}{15 \pi}
\int 
k^2 \alpha_k^2
    \left(n_k + \frac{1}{2}\right)
    \sin (2\varepsilon_k t)
    \D k.
\end{equation}
Importantly, the signal is insensitive to the sign of $\alpha_k$: the permittivity oscillations begin with an increase along the pump polarization. For short pulses, the response is linear in pump-pulse energy, consistent with the experiment.

After a few picoseconds, the oscillations at different frequencies dephase, leaving only stationary-phase components. The helium dispersion curve has three such points: the maxon maximum (at $k_M=1.1\,\text{\AA}^{-1}$), the roton minimum (at $k_R=1.9\,\text{\AA}^{-1}$), and Pitaevskii's plateau near the dispersion termination point ($k_P=3.6\,\text{\AA}^{-1}$). In all three cases, Bose-Einstein occupation numbers are negligible, $n_k \ll 1/2$, at all temperatures below the lambda point, so the signal is dominated by zero-point vibrations. In the stationary-phase approximation the energy is expanded as $\varepsilon_k = \Delta + (k-k_0)^2/2m$ near each non-degenerate extremum, thus reducing the problem to the evaluation of Fresnel integrals. The contributions of maxons and rotons obtained in this way are:
\begin{align}
\epsilon_M^{mn}&=
\frac{2 E_m E_n k_M^2 \alpha_M^2 \tau |m_M|^{1/2}}{15\pi^{1/2} t^{1/2}}
\sin\left(2\Delta_M t - \frac{\pi}{4}\right),
\\
\epsilon_R^{mn}&=
\frac{2 E_m E_n k_R^2 \alpha_R^2 \tau m_R^{1/2}}{15\pi^{1/2} t^{1/2}}
\sin\left(2\Delta_R t + \frac{\pi}{4}\right),
\end{align}
where $\Delta_M = 13.9\,\text{K}$, and $\Delta_R = 8.6\,\text{K}$ are the maxon and roton energies, and $m_M=-0.55\,m_\text{He}$ and $m_R=0.14\,m_\text{He}$ are the corresponding masses~\cite{data}. Note the different phase shifts of the maxon and roton contributions due to the different signs of the masses.

Pitaevskii's plateau is a degenerate stationary point of infinite order~\cite{LL9}. Its contribution can be estimated as
\begin{equation}
\epsilon_P^{mn} \sim
\frac{2E_m E_n k_P^3 \alpha_P^2 \tau}{105 \pi}
\sin (4 \Delta_R t).
\end{equation}

\section{Discussion}
The observed opposite-momentum two-mode squeezing is a direct consequence of the parametric laser coupling~\eqref{h-definition}. The angular structure of the squeezed state, \eqref{alpha-general}, corresponds to $l=2$, $m=0$, with quantization axis along the linear pump polarization. With circular pump polarization, one expects a different $d$-wave squeezed state whose quantization axis is along the propagation direction; oscillations in this state should be visible with a probe perpendicular to the pump. Preliminary experiments by the Lancaster University team~\cite{lancaster_private} appear to support this prediction.

Careful analysis of experimental data confirms the presence of weak contributions from maxons and Pitaevskii's plateau~\cite{lancaster_private,milner_private}, alongside the primary roton signal. The exact cause of such small amplitudes of these contributions remains unclear.

In the present work, we considered free quasiparticles. Interactions between them lead to two important effects. First, exponential signal decay due to the finite lifetime of quasiparticle pairs masks the $\propto t^{-1/2}$ amplitude dependence (which is a hallmark of the dephasing frequency selection mechanism). Note that the pairs have no net momentum, so the decay is not expected to disappear even at zero temperature, despite the infinite lifetime of individual quasiparticles.

A related point concerns the dynamics of the scalar part of permittivity, which is less sensitive to zero-temperature decay. The corresponding isotropic component of quasiparticle polarizability cannot be estimated by the same approach as \eqref{alpha} and is not relevant for the birefringence signal. Nevertheless, oscillations of the squeezed $s$-state contribute to the local refractive-index modulation and should therefore be detectable by beam deflection.

Second, the interactions offset the oscillation frequencies; the experimentally observed frequency of the roton contribution is indeed slightly lower than $2\Delta_R$~\cite{exp}. This is consistent with known negative roton-roton interaction in the zero-momentum $d$-wave channel. This frequency offset is probably responsible for the observed phase shift pattern~\cite{milner_private}, which is more complicated than predicted by the stationary-phase approximation and requires further investigation.

For a short pump pulse, the birefringence response \eqref{result} is linear in its duration $\tau$. To describe longer pulses, full expressions for the operators \eqref{operators} should be used, which are given by \eqref{operators-exact} in the Appendix.

\section*{Acknowledgments}
I thank M.\,Elbaum, M.\,Feigelman, S.\,Kafanov, V.\,Marchenko, R.\,Mikhaylovskiy, V.\,Milner, and E.\,Surovtsev for useful discussions.

\section*{Appendix}
Evaluation of the expressions \eqref{operators} with quadratic Hamiltonian $\hat{h}$ is a typical problem in a two-mode squeezing setup. It can be solved by integrating the Heisenberg system of equations of motion for the coupled operators:
\begin{equation*}
\begin{aligned}
\hat{A}_+ &=  \E^{\I \hat{h} \tau}
                        \,\a{+}\,
                    \E^{-\I \hat{h} \tau},\\
\hat{A}^\dagger_- &=  \E^{\I \hat{h} \tau}
                        \,\aa{-}\,
                    \E^{-\I \hat{h} \tau}.
\end{aligned}
\end{equation*}
It follows that
\begin{equation}
\label{system}
\begin{aligned}
\frac{\D\hat{A}_+}{\D\tau} &=
    -\I\left((\varepsilon + \zeta)\hat{A}_+ + \zeta\hat{A}^\dagger_- \right),\\
\frac{\D\hat{A}^\dagger_-}{\D\tau} &=
    \hphantom{-}\I\left((\varepsilon + \zeta)\hat{A}^\dagger_- +\zeta\hat{A}_+ \right),
\end{aligned}
\end{equation}
where the commutation relation with Hamiltonian \eqref{h-definition}
\begin{equation*}
[\hat{h}, \a{+}] = 
-\left(\varepsilon+\zeta\right) \a{+} -\zeta \aa{-}
\end{equation*}
is taken into account. One can easily verify that the following expressions satisfy the initial conditions
\begin{equation*}
\left.\hat{A}_+\right|_{\tau=0} = \a{+}, \quad \left.\hat{A}^\dagger_-\right|_{\tau=0} = \aa{-}
\end{equation*}
and solve the system \eqref{system}:
\begin{equation}
\label{operators-exact}
\begin{aligned}
\hat{A}_+ &= \left(\cos\Omega\tau -\I\frac{\varepsilon+\zeta}{\Omega}\sin\Omega\tau\right)\a{+}-\I\frac{\zeta\sin\Omega\tau}{\Omega} \aa{-},\\
\hat{A}^\dagger_- &= \left(\cos\Omega\tau +\I\frac{\varepsilon+\zeta}{\Omega}\sin\Omega\tau\right)\aa{-}+\I\frac{\zeta\sin\Omega\tau}{\Omega} \a{+},
\end{aligned}
\end{equation}
where $\Omega^2 = \varepsilon^2 + 2\varepsilon\zeta$. In the limit $\Omega\tau\ll 1$, we recover~\eqref{operators-approx}.

\end{document}